\documentclass[useAMS,usenatbib]{mn2e}
\pdfoutput=1
\usepackage[british]{babel}
\usepackage{graphicx}
\usepackage{epstopdf}
\usepackage{color}
\usepackage{amsmath}

\newcommand{\rxte}{{\it RXTE}}

\newcommand{\rxtepca}{{\it RXTE}/PCA}
\newcommand{\rxtehexte}{{\it RXTE}/HEXTE}

\newcommand{\isis}{{\sc isis}}

\newcommand{\heasoft}{{\sc heasoft}}

\newcommand{\fastica}{{\sc fastica}}
\newcommand{\nmf}{{\sc nmf}}

\newcommand{\diskbb}{{\sc diskbb}}
\newcommand{\phabs}{{\sc phabs}}
\newcommand{\cutoffpl}{{\sc cutoffpl}}

\title[Unsupervised spectral decomposition of XRBs]{Unsupervised spectral decomposition of X-ray binaries with application to GX 339--4}

\author[K. I. I. Koljonen]
{K.~I.~I.~Koljonen$^{1,2}$\thanks{email: karri.koljonen@nyu.edu}
\\
$^{1}$Aalto University Mets\"ahovi Radio Observatory, Mets\"ahovintie 114, FIN-02540 Kylm\"al\"a, Finland \\
$^{2}$New York University Abu Dhabi, PO Box 129188, Abu Dhabi, UAE}

\begin{document}

\pagerange{\pageref{firstpage}--\pageref{lastpage}}
\pubyear{2014}

\maketitle

\label{firstpage}

\begin{abstract} 

In this paper, we explore unsupervised spectral decomposition methods for distinguishing the effect of different spectral components for a set of consecutive spectra from an X-ray binary. We use well-established linear methods for the decomposition, namely principal component analysis, independent component analysis and non-negative matrix factorisation (NMF). Applying these methods to a simulated dataset consisting of a variable multicolour disc black body and a cutoff power law, we find that NMF outperforms the other two methods in distinguishing the spectral components. In addition, due the non-negative nature of NMF, the resulting components may be fitted separately, revealing the evolution of individual parameters. To test the NMF method on a real source, we analyse data from the low-mass X-ray binary GX 339--4 and found the results to match those of previous studies. In addition, we found the inner radius of the accretion disc to be located at the innermost stable circular orbit in the intermediate state right after the outburst peak. This study shows that using unsupervised spectral decomposition methods results in detecting the separate component fluxes down to low flux levels. Also, these methods provide an alternative way of detecting the spectral components without performing actual spectral fitting, which may prove to be practical when dealing with large datasets.      

\end{abstract}


\section{Introduction} \label{introduction}

Scientific consensus dictates that the X-ray spectra of black hole X-ray binaries (XRBs) and active galactic nuclei (AGNs) are modelled by three spectral components: a multicolour black body component representing the emission from the accretion disc in the soft X-rays, a power law with a cutoff or no cutoff representing the thermal/non-thermal Comptonisation of soft photons from a population of hot electrons in the hard X-rays, and a reflection component representing the reprocessed emission of hard X-rays from the accretion disc, i.e. photoelectrically absorbed or Compton-scattered emission, in the intermediate energies, that most prominently manifests itself in the form of an iron line. However, the magnitude of these components remains unknown at a given time. For example, the magnitude of the soft component in the hard state is usually difficult to determine. However, knowledge of the accretion disc geometry in the hard state, in particular the location of the inner edge of the accretion disc, is critical for the properties and/or the validity of all accretion disc models. It is especially important for radiatively inefficient models such as advection dominated accretion flow \citep[ADAF,][]{ichimaru,narayan94}. The quiescent properties of black hole transients are often cited as evidence for ADAFs and black hole event horizons \citep{narayan97,garcia,mcclintock}. However, for the ADAF model to be relevant, the inner edge of the accretion disc must increase by orders of magnitude between the outburst and quiescent states. Therefore, it is important to determine the magnitude of the accretion disc and reflection components in the hard state -- a difficult task due to the low flux compared to the Comptonisation component. In fact, even the physical mechanism for the X-ray emission in the hard state is open to question, with possible contributions from direct synchrotron emission from the jet \citep{falcke,markoff01,russell} or synchrotron self-Compton radiation from the base of the jet \citep{markoff05}. This kind of spectral degeneracy is especially true for the Galactic XRBs and a striking example can be seen e.g. in \citet{nowak11}, where three very different models are fitted equally well to the same data set of Cyg X-1, despite the excellent quality of the data that was obtained by all the X-ray satellites in orbit at the time. In addition, the magnitudes of different components in the intermediate X-ray states are difficult to determine as all the components are luminous. Extra complications in the interpretation of X-ray spectra are caused by the attenuation of interstellar matter in Galactic XRBs and the so-called warm absorbers (possibly associated with the winds of the accretion discs) in AGNs. Thus, modelling the X-ray spectra of accreting black holes often leads to a problem of degeneracy, i.e. multiple distinct models fit the observed data equally well. Even if an apparently good fit is obtained between the data and the model, it does not necessarily imply a match between theory and physical reality.  

In this paper we demonstrate the use of linear unsupervised decomposition methods in separating a set of time series of X-ray spectra into subcomponents corresponding to distinct spectral components from the disc and the population of hot electrons. In Section \ref{spectral_decomposition} we will compare three different decomposition methods, namely principal component analysis (PCA), independent component analysis (ICA) and non-negative matrix factorisation (NMF), using simulated spectra mimicking the spectral behaviour from a typical XRB. This analysis provides a better estimate of the X-ray continuum models required to fit the X-ray spectra in XRBs by taking into account also the spectral variability in addition to the fitting of the time-averaged spectra. In Section \ref{discussion} we apply the NMF to a set of spectra from the stellar mass black hole XRB GX 339--4. With a sufficiently long set of observations throughout the hardness-intensity diagram (HID) from GX 339$-$4, we study how spectral decomposition reveals the change of spectral components in the HID. In Section \ref{conclusions} we present the conclusions of this paper and make suggestions for the future use of the spectral decomposition methods in the context of XRBs.  

\section{Unsupervised spectral decomposition} \label{spectral_decomposition}

X-ray spectra can be decomposed via matrix factorisation techniques. Generally, this problem falls into the category of blind source separation (BSS), in which a set of source signals is estimated from a set of mixed signals with no information on the source signals or the mixing process. In this case the data matrix is a collection of discrete X-ray spectra, $X_{ji}$, consisting of flux values with energy $j$ measured at time $i$. This matrix is taken to consist of a mix of a few, up to $k$, separate source signals $S_{ki}$ weighted over the energy bands by a weight matrix $W_{jk}$. Essentially, the columns of the weight matrix can be thought to represent a constant spectral component that has a variable amplitude dictated by the rows of the source signal matrix for each $k$. In practice, $k$ is much smaller than $i$ or $j$, as the intention is to describe the information in $X$ concisely. $S$ and $W$ are subsequently estimated in such a way that $X \approx WS$, i.e.    

\begin{equation} \label{eq1}
X_{ji} \approx \sum_{k}W_{jk}S_{ki}.      
\end{equation}
  
The BSS problem is very underdetermined since multiple equally valid solutions exists. However, by placing clever requirements on the factorisation one can reduce the number of possible solutions and obtain meaningful results. PCA, for example, assumes that the source signals are minimally correlated to each other and ICA assumes that the source signals are maximally independent of each other. A slightly different approach is introduced in NMF which imposes structural constraints, i.e. the non-negativity of the constituent matrices $S$ and $W$. In the following we create a simulated dataset that mimics the spectral effects that are present in the usual X-ray data of XRBs and tackle it using the methods outlined above.         

\subsection{Simulations}

\begin{figure}
\begin{center}
\includegraphics[width=0.5\textwidth]{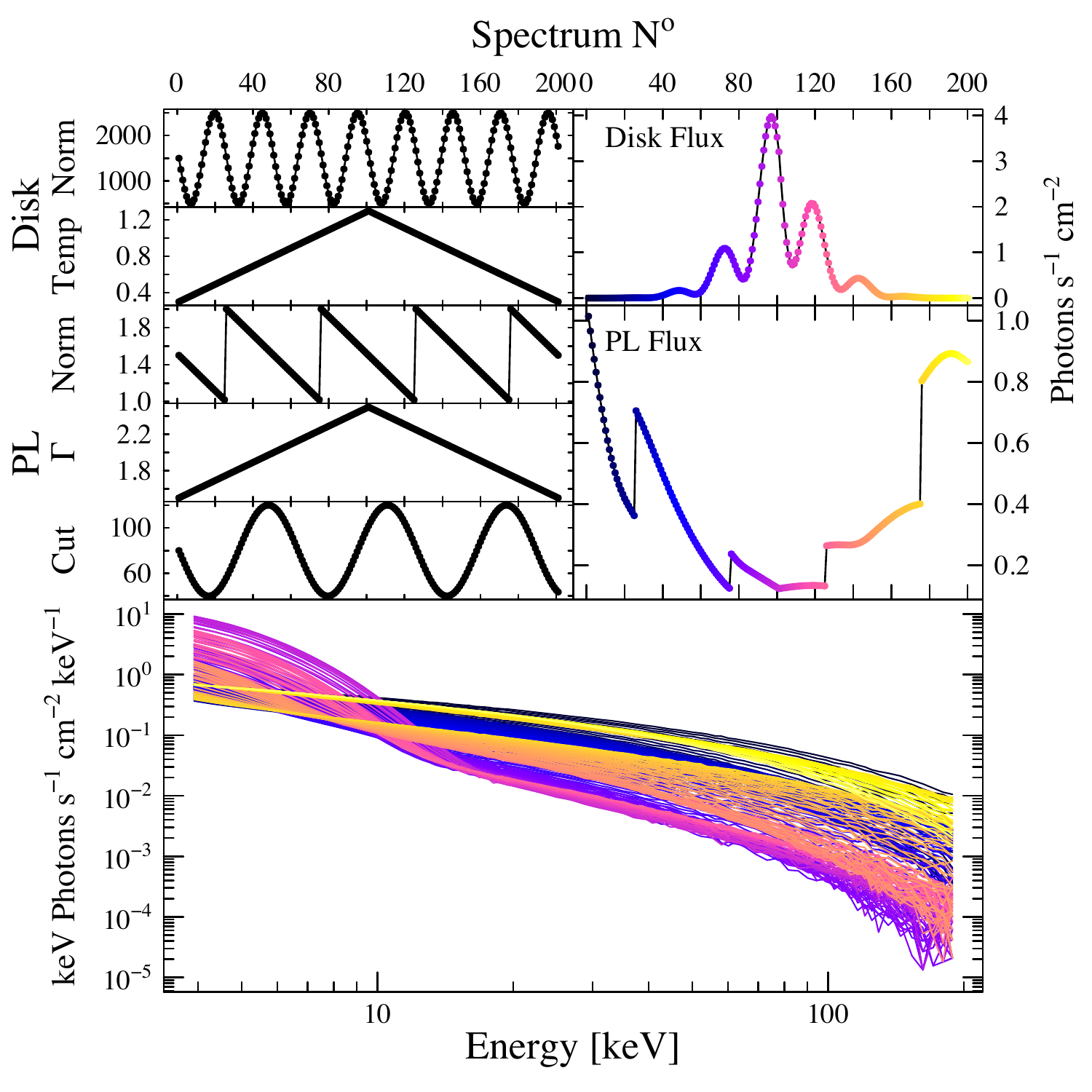}
\end{center}
\vspace{-12pt}
\caption{The parameters, fluxes and resulting X-ray spectra of the simulated data for testing the different matrix factorisation methods. The five panels in the upper left side of the figure show the variability of the parameters of individual spectral components: top two for multicolour disc black body (sine wave for the disc normalisation and sloped line for the disc temperature), and bottom three for cutoff power law (saw wave for power law normalisation, sloped line for the spectral index, and sine wave for the cutoff energy). The two panels in the upper right side of the figure show the corresponding flux values from the disc and power law components. The bottom panel shows the resulting X-ray spectra faked using the models described above.} \label{sim}
\end{figure}

In order to gauge the appropriateness of a decomposition method (PCA/ICA/NMF) we created a simulated dataset that mimics the spectral effects that are present in the usual X-ray data of XRBs: absorbed (\phabs) disc black body spectra (\diskbb) with changing normalisation and temperature values and cutoff power law spectra (\cutoffpl) with changing normalisation, spectral index and cutoff values. For `spectral pathways' we use distinguishable functions such as sine and saw waves. In addition, we vary the disc temperature and the power law index so as to produce soft and hard X-ray states. We use \isis\/ \citep{isis} to fake 200 spectra with an exposure of 5 ks, varying the black body normalisation sinusoidally from 500 to 2500, the black body temperature linearly from 0.3 keV to 1.3 keV and back, the power law normalisation as a saw wave from 1.0 to 2.0, power law index linearly from 1.5 to 2.5 and back, and a power law cutoff sinusoidally from 40 keV to 120 keV (Fig. \ref{sim}). We used the \rxte\/ response function from GX 339--4 pointing 70110--01--04--00 and then unfolded the spectra using \isis\/ to flux units (keV photons s$^{-1}$ cm$^{-2}$ keV$^{-1}$). We would like to note that by using ``flux-corrected'' spectra, response matrix features are introduced in addition to the physical processes, which could have an effect in the low flux regimes.

As PCA, ICA and NMF are all methods that rely on \textit{linear} decomposition, they are suited best for finding the variability of the normalisations of spectral components, and can behave erratically when dealing with more complicated effects such as changing optical depth and electron temperature that varies the cutoff and spectral slope of the X-ray spectra. When presented with data that varies non-linearly, linear methods can, however, be used as the non-linearity can be approximated as a collection of several linear components. Thus, we do not expect the decomposition to return the variability of the parameters themselves, but rather the fluxes of the spectral components that are collectively produced by the varying parameters. The quality of the decomposition is then measured by how well they are able to reproduce the disc and power law fluxes. In the following we review the different spectral decomposition methods and how they are able to reproduce the original fluxes of the simulated spectral components. 

\subsection{Principal component analysis}

PCA \citep[review e.g.][]{jolliffe} is one of the standard tools of time series analysis that has been used in astronomy mainly for stellar spectral classification \citep[e.g.][]{whitney}, galaxy spectral classification \citep[e.g.][]{connolly}, and quasar spectral classification \citep[e.g.][]{francis}. For studying variable X-ray spectra PCA has been used in AGNs e.g. by \citet{vaughan} and \citet{parker}, and in XRBs by \citet{malzac} and \citet{koljonen}. However, one of the drawbacks in using PCA for decomposing spectra is its tendency to cancel components as it works with the mean-subtracted spectra. This results in components that have either too exaggerated or diminished an impact on the X-ray spectra depending on the values of their normalisation. Additionally, if the eigenvectors of the principal components have both positive and negative values, it results in pivoting behaviour of the spectra, i.e. when the normalisation of that component increases it increases the impact of the component in the positive part and decreases the impact in the negative part of the eigenvector. However, this turns out to be a good proxy for the power law index \citep{parker}. 

We have used a singular value decomposition (SVD) in calculating the components of individual, mean-subtracted spectra that comprise $X$ on the left-hand side of Eq. \ref{eq1}. More detailed description of the method can be found in e.g. \citet{malzac}, \citet{koljonen} and \citet{parker}.   

\subsubsection{Choosing the degree of the factorisation} \label{degree}

\begin{figure}
\begin{center}
\includegraphics[width=0.5\textwidth]{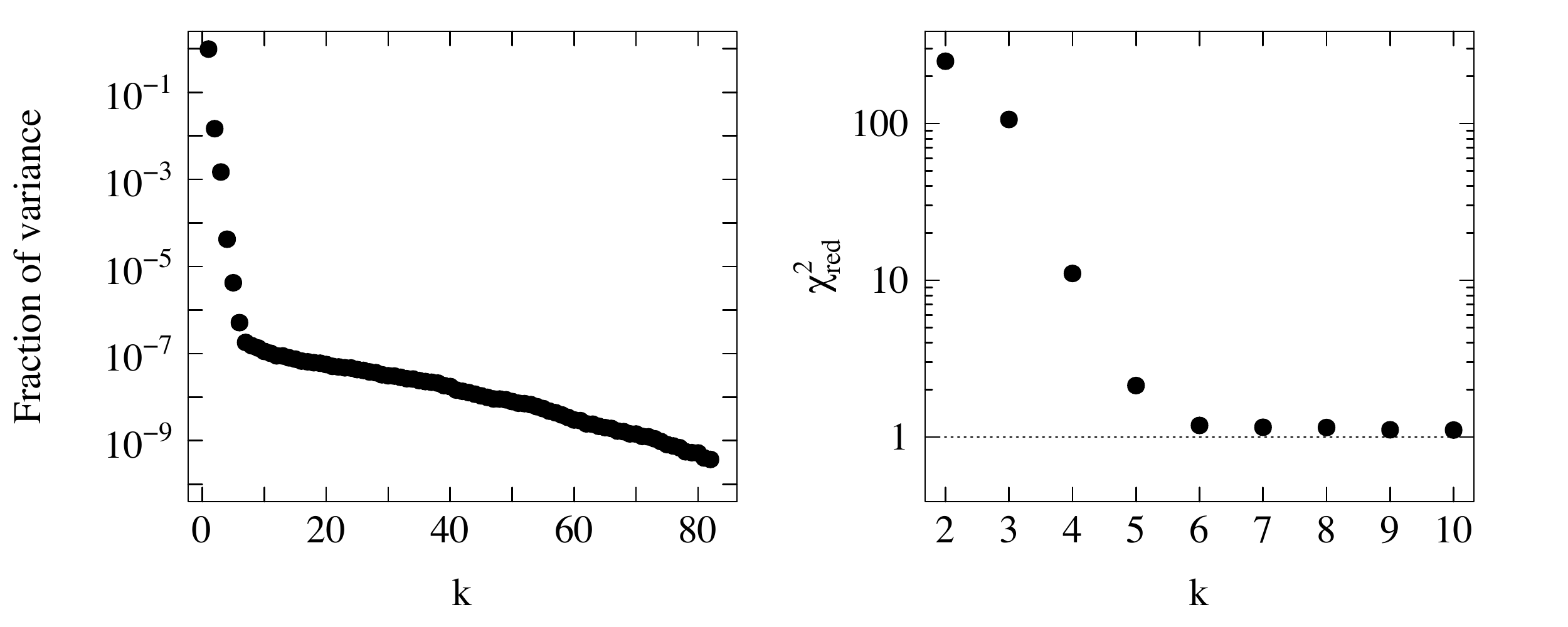}
\end{center}
\vspace{-12pt}
\caption{Determining the degree of factorisation for PCA. \textit{Left:} the log-eigenvalue (LEV) diagram of the simulated dataset. \textit{Right:} the $\chi^{2}$-diagram of the simulated dataset. After $k=6$ the LEV settles down to a straight line indicating the start of the noise. Likewise, in the $\chi^{2}$-diagram $k$ achieves the value where further adding the number of components decreases the reduced $\chi^{2}$ value only by a small amount after $k=6$.} \label{pcalev}
\end{figure}

\begin{figure}
\begin{center}
\includegraphics[width=0.5\textwidth]{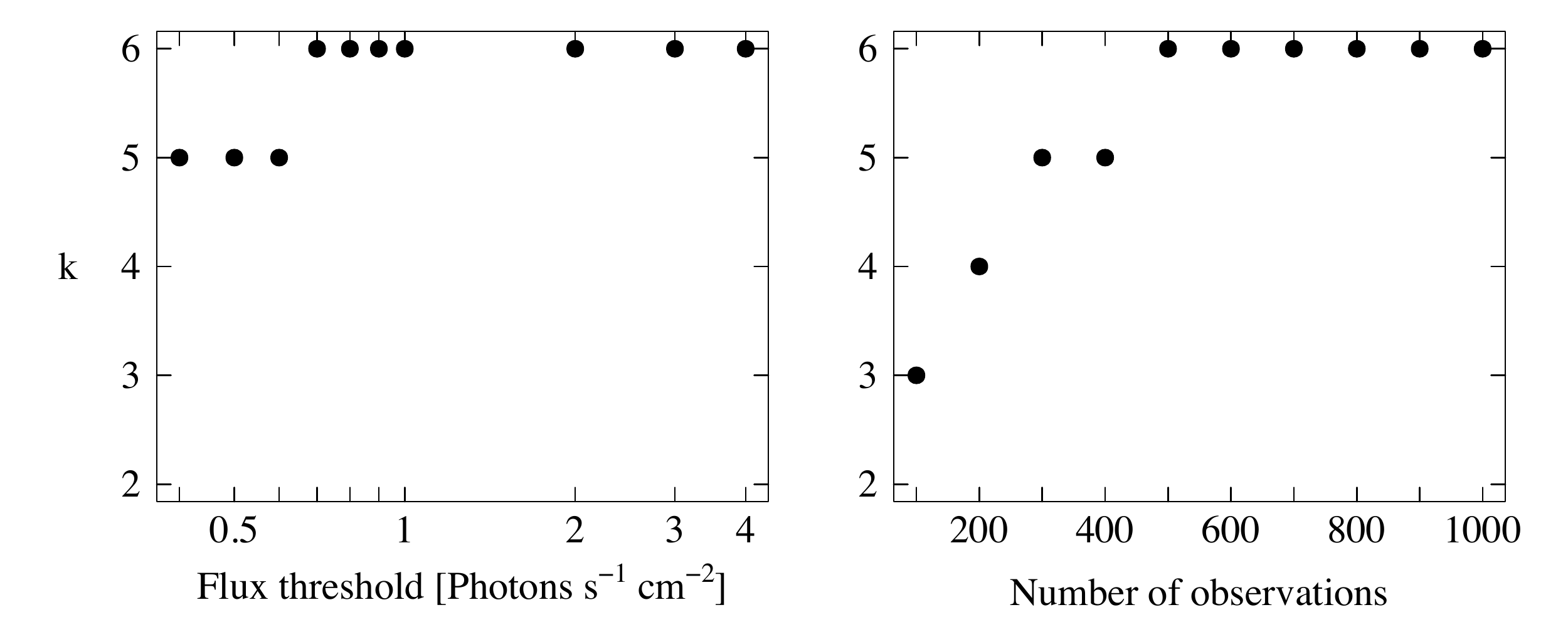}
\end{center}
\vspace{-12pt}
\caption{The effect of sampling the data to the number of significant factorised components above the noise level in the $\chi^{2}$-diagram. \textit{Left:} The number of $k$ components above the noise level when the data is selected to be below a certain flux threshold. \textit{Right:} The number of $k$ components above the noise level when the data is selected to span a certain number of observations from the beginning of the simulation.} \label{clips}
\end{figure}

In general, we would expect the quality of the factorisation, i.e. its similarity with the original data, to be an increasing function of the degree of the factorisation $k$. A ``good'' value of $k$ would not be a local maximum for quality, but instead a point where the response of the quality to $k$ changes from being steep to shallow (i. e., a ``good'' value of $k$ provides a substantially better approximation than nearby smaller values, but only a slightly worse approximation than nearby larger values). One method of choosing the degree of the factorisation in PCA is the log-eigenvalue (LEV) diagram \citep[e.g.][]{jolliffe,koljonen}. In the LEV diagram significant components can be distinguished from the noise by their deviation from the geometrical progression, i.e. straight line in the diagram. In addition, we devise a method similar to LEV diagram to measure the quality of the factorised spectra using a median of reduced $\chi^{2}$ -values of the resulting factorisation when compared to the simulated dataset. This method calculates the reduced $\chi^{2}$-values between each individual spectrum from the factorisation $WS$, on the right-hand side of Eq. \ref{eq1}, with different degrees of factorisation and the spectra from the simulated dataset with associated errors and takes the median of these values, thus producing a quality measure of how well the factorisation with degree $k$ fits in to the data and reducing the number of components to those that vary above the noise level. As mentioned above the chosen degree of factorisation should be a point where the median of reduced $\chi^{2}$ -value changes from being steep to shallow. In addition, the value should be close to 1 so as to portray faithfully the original spectra. Let us call this method the $\chi^{2}$-diagram and it can be formulated as follows:

\begin{equation}
\chi^{2}_{red}(k) = \mathrm{Mdn} \Bigg\{ \frac{\sum_{i} [(X_{ji}-\sum_{k}W_{jk}S_{ki})/\sigma_{ji}]^{2}}{\mathrm{max}(j)-k} \Bigg\}.
\end{equation}

The $\chi^{2}$-diagram will be used for ICA and NMF results as well so as to make the comparison between different methods easier. This is because the LEV diagram cannot be used for ICA and NMF as in these methods the number of components of the factorisation has to be determined before starting the analysis itself. 

Fig. \ref{pcalev} shows the LEV diagram and the $\chi^{2}$-diagram for the simulated dataset with different degrees of factorisation. The LEV diagram in Fig. \ref{pcalev} indicates the start of the noise after $k=6$, thus indicating that six components is sufficient to explain the simulated X-ray spectra. Likewise, in the $\chi^{2}$-diagram after $k=6$ there appears to be no improvement to the $\chi^{2}_{red}$ value by adding more components, and thus it is sufficient to take six components into account. This also demonstrates that the $\chi^{2}$-diagram leads to the same conclusion as the LEV diagram. Based on these methods it suffices to take into account six principal components for explaining the simulated X-ray spectra.   

We also studied how the number of $k$ above the noise level changes in the $\chi^{2}$-diagram on a flux-limited or time-limited sample. In this case, we use the same simulation setup, models, and parameter ranges, but increase the number of spectra five-fold to increase resolution. Fig. \ref{clips} shows how $k$ changes if the sample is restricted below some flux threshold (left panel), or if the sample is restricted to span a number of observations from the beginning of the simulation (right panel). To obtain $k=6$, we see that use of the whole dataset is not necessary, although it is still a substantial amount (the flux level 0.7 photons s$^{-1}$ cm$^{-2}$ roughly divides the data in two equal sets). However, even in smaller samples the number of significant components is only slightly smaller.

\begin{figure}
\begin{center}
\includegraphics[width=0.5\textwidth]{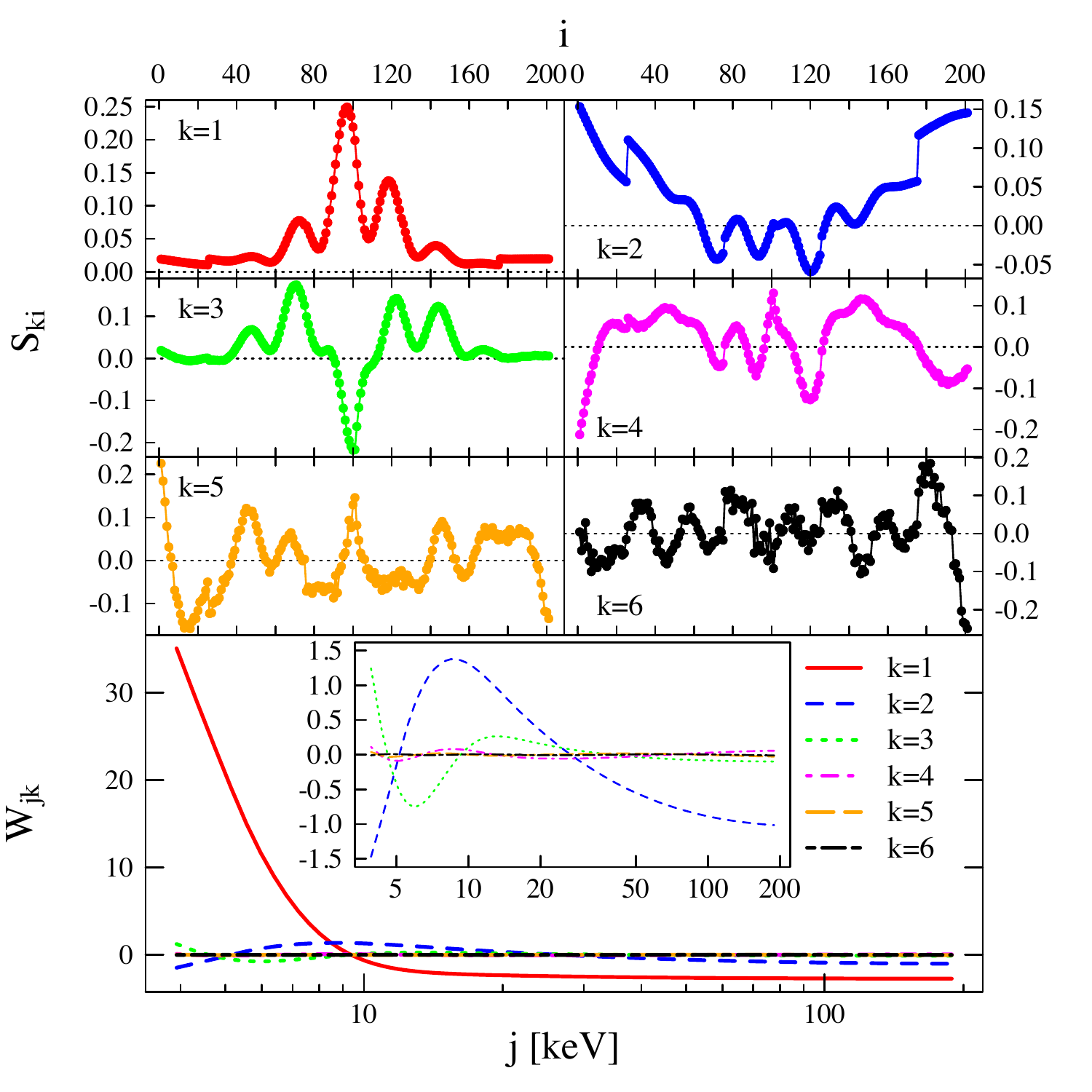}
\end{center}
\vspace{-12pt}
\caption{The six components from the PCA sufficient to explain most of the variability in the X-ray spectra of the simulated dataset. The top six panels show the source signals $S_{ki}$ of the principal components and the bottom panel shows the weights $W_{jk}$ of the principal components with the inset showing the last five with a zoomed range for the y-axis.} \label{pca}
\end{figure}

\subsubsection{Simulated data}

Fig. \ref{pca} shows the six principal components; the source signals $S_{ki}$ and weights $W_{jk}$ across the energies, derived from the simulated dataset. It is clear that the first component ($k=1$) corresponds to the variations in the disc component, and the second component ($k=2$) correspond to the variations in the power law component. However, it is not clear how the remaining components contribute to the disc and power law components and they likely produce random negative and positive corrections to the disc and/or power law components. A comparison of the first two components with the disc and power law fluxes from the simulation is shown in Section \ref{quality}.  

\subsection{Independent component analysis}

ICA \citep[for a review see e.g.][]{hyvarinen01} is a linear solution to BSS similar to PCA, but instead of uncorrelatedness and orthogonality ICA relies on statistical independency of the constituent components. Similarly to PCA, ICA has been used in astronomy mainly for galaxy spectral classification \citep[e.g.][]{lu,allen}. Statistical independency is a more strict rule than uncorrelatedness, as the uncorrelatedness is implied in independency but not vice versa. ICA relies on the central limit theorem which states that the mean of random processes will always approach gaussian distribution. Thus, ICA searches for non-gaussian components. ICA suffers from the same problems in spectral decomposition as PCA, as it also works with the mean-substracted spectra resulting in positive and negative weights and pivoting source spectra. 

We have used a \fastica\/ algorithm \citep{hyvarinen99} in calculating the components of individual, pre-whitened spectra that comprise $X$, on the left-hand side of Eq. \ref{eq1} using negative entropy to search for the non-gaussian components.       

\subsubsection{Choosing the degree of the factorisation}

\begin{figure}
\begin{center}
\includegraphics[width=0.5\textwidth]{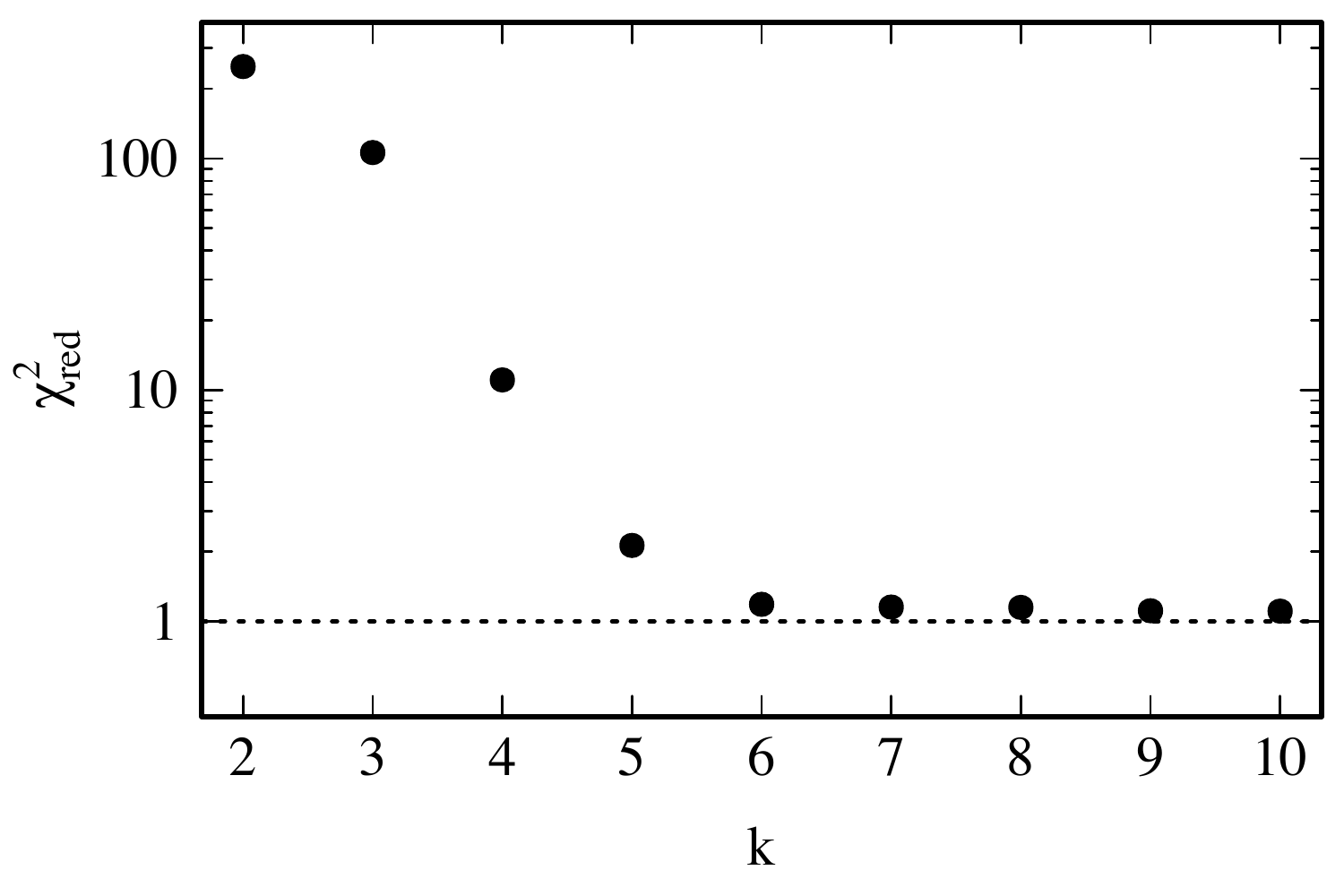}
\end{center}
\vspace{-12pt}
\caption{Determining the degree of factorisation for ICA. The figure shows the $\chi^{2}$-diagram of the simulated dataset. After $k=6$ the $\chi^{2}$-diagram achieves the value where further adding the number of components decreases the reduced $\chi^{2}$ value only by a small amount, thus it is sufficient to take six components into account for the matrix factorisation.} \label{icachi}
\end{figure}

ICA has the disadvantage to PCA that it does not arrange the components via their fraction of variance. Additionally, the factorisation degree has to be determined before running the algorithm. As in PCA, we use the same method to determine the quality of the factorisation using the $\chi^2$-diagram as explained in Section \ref{degree}. Fig. \ref{icachi} shows the $\chi^2$-diagram for different degrees of factorisation. Similar to PCA results, after $k=6$ there appears to be no improvement to the $\chi^{2}_{red}$ value by adding more components, and thus it is enough to take six components into account for the ICA analysis.  

\begin{figure}
\begin{center}
\includegraphics[width=0.5\textwidth]{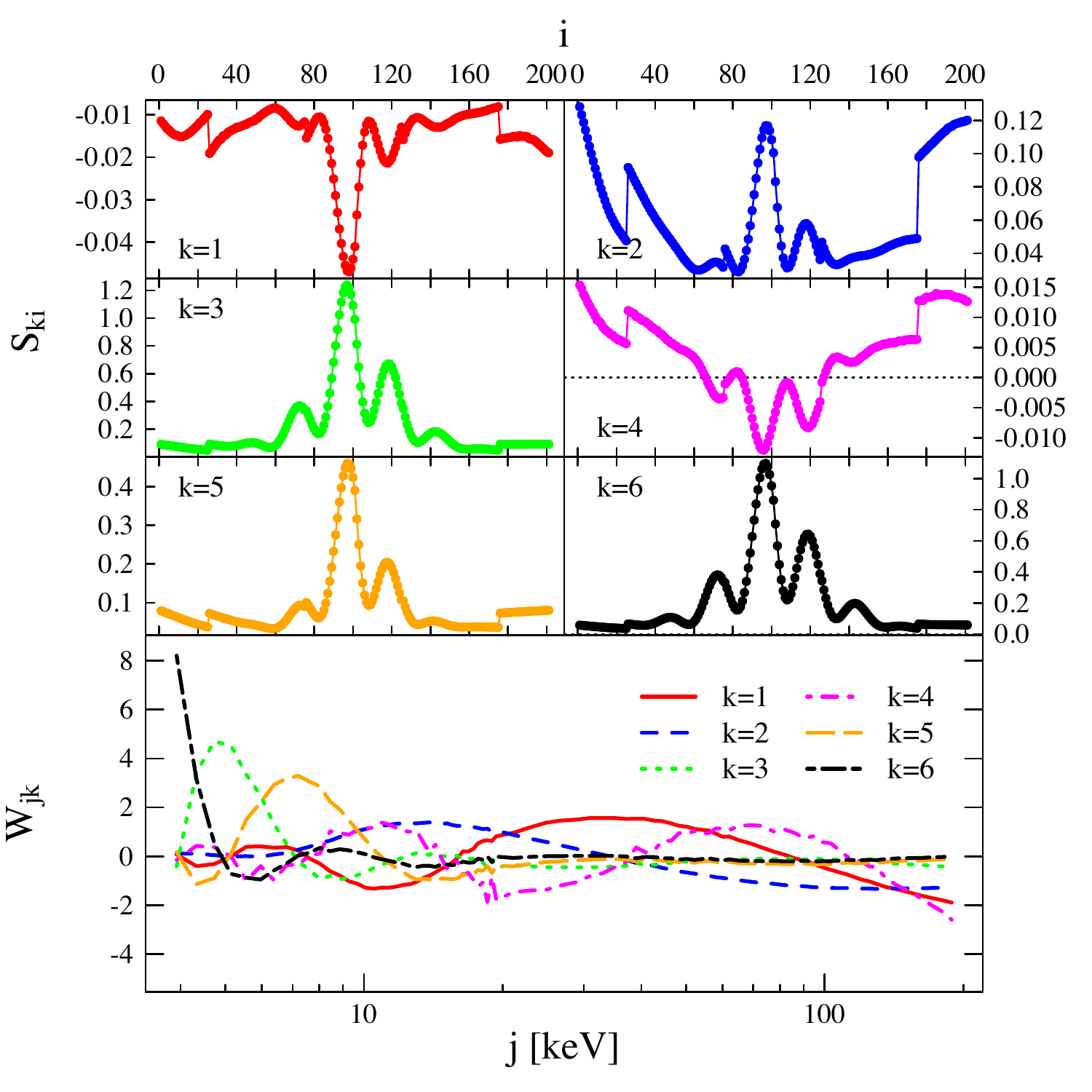}
\end{center}
\vspace{-12pt}
\caption{The six components from the independent component analysis sufficient to explain most of the variability and X-ray spectra of the simulated dataset. The top six panels show the source signals $S_{ki}$ and the bottom panel shows the weights $W_{jk}$ of the independent components.} \label{ica}
\end{figure}

\subsubsection{Simulated data}

Fig. \ref{ica} shows the six independent components, divided into the source signals and weights as previously, derived from the simulated dataset. Unlike in PCA, all the components seem to be relevant in either the disc and/or power law component. Based on the energy range of the weights and the simulated fluxes in Fig. \ref{sim} it appears that the $k=3,5,6$ independent components represent the disc and $k=1,2,4$ independent components represent the power law spectral components. However, it can be seen that when the power law flux is low the disc flux leaks to all independent components. A comparison of the disc and power law components as represented by the respective independent components with the disc and power law fluxes from the simulation is shown in Section \ref{quality}.  

\subsection{Non-negative matrix factorisation} \label{nmf_sec} 

NMF \citep{paatero,lee99} provides an exciting alternative to traditional dimensional-reduction methods. In NMF, samples are represented by non-negative combinations of canonical components. The structure found by NMF methods is thus often very different from, and more intuitive to interpret than, that of more traditional eigenvector-based methods, such as PCA. By constructing samples as positive combinations, NMF also has the potential to disentangle canonical components which often overlap to create particular community samples. The price of this more intuitive decomposition is that the factorisation is approximate, i.e. not unique as in PCA, and components depend on the dimension of the decomposition, requiring greater care in the interpretation.         

In NMF the matrices $W$ and $S$ are found by minimising a cost function under the constraint that they must be non-negative. Such a cost function can be constructed using some distance measure $D$ between $X$ and $WS$. In this paper, we use the generalised Kullback-Leibler (KL) divergence, 

\begin{equation}
D_{KL}(X||M) = \sum_{ji} \Big( X_{ji} \mathrm{log} \frac{X_{ji}}{M_{ji}} - X_{ji} + M_{ji} \Big), 
\label{cost}
\end{equation}

where $M_{ji} = \sum_{k}W_{jk}S_{ki}$. The KL divergence measures the information lost when $M$ is used to approximate $X$, and thus minimising it will result in maximising the information of $X$ in $M$. Additionally, it has been shown in \citet{sajda} that minimising the above cost function is equivalent to maximising the likelihood of generating $X$ from $M$ when $X$ is assumed to be Poisson distributed with mean $M$. Therefore, the KL divergence is appropriate cost function to use with X-ray counting data.   

In this paper we use the \nmf\/ package \citep{gaujoux} that calculates the standard NMF \citep{brunet} by picking random starting values for $W$ and $S$ from a uniform distribution $[0,\rm{max}(X)]$ and then updating iteratively 10000 times to find a local minimum of the cost function with a multiplicative rule from \citet{lee01}:  

\begin{equation}
W_{jk} \leftarrow W_{jk} \Big( \sum_{i} \frac{S_{ki} X_{ji}}{M_{ji}} \Big) \Big( \sum_{i} S_{ki} \Big)^{-1}, 
\end{equation} 

\begin{equation}
S_{ki} \leftarrow S_{ki} \Big( \sum_{j} \frac{W_{jk} X_{ji}}{M_{ji}} \Big) \Big( \sum_{j} W_{jk} \Big)^{-1}. 
\end{equation} 

To ensure that the algorithm does not get stuck in a local minimum we repeated the minimisation process for 50 different starting points (30--50 starting points is considered sufficient in \citealt{hutchins} and \citealt{brunet} to get robust estimate of the factorisation degree $k$) for each value of $k$ and 300 different starting points for the chosen degree of factorisation, and used the factorisation which minimised the cost function in Eq. \ref{cost}.

\subsubsection{Choosing the degree of the factorisation} \label{nmfdegree}

\begin{figure}
\begin{center}
\includegraphics[width=0.5\textwidth]{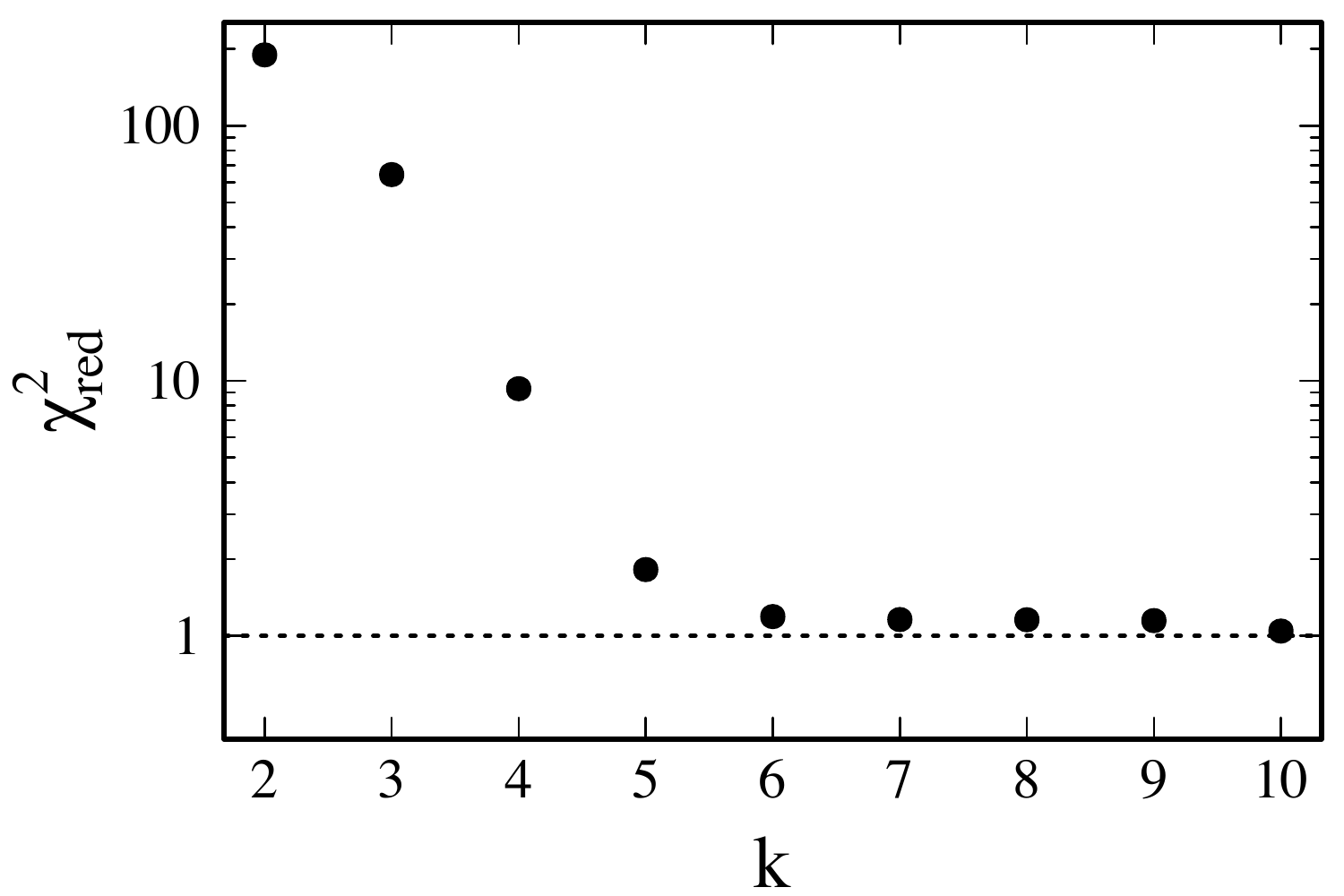}
\end{center}
\vspace{-12pt}
\caption{Determining the degree of factorisation for NMF. The figure shows the $\chi^{2}$-diagram of the simulated dataset. After $k=6$ the $\chi^{2}$-diagram achieves the value where further adding the number of components decreases the reduced $\chi^{2}$ value only by a small amount.} \label{nmfchi}
\end{figure}

NMF is not a hierarchical method; each component depends on the choice of degree $k$ and thus this choice should be made with care. Similar to PCA and ICA above, we use the $\chi^{2}$-diagram as explained in Section \ref{degree}. Fig. \ref{nmfchi} shows the $\chi^2$-diagram for different degrees of factorisation. After $k=6$ there seems to be no improvements to the $\chi^{2}_{red}$ value by adding more components, and thus it is enough to take six components into account for the NMF analysis.      

\subsubsection{Simulated data}

\begin{figure}
\begin{center}
\includegraphics[width=0.5\textwidth]{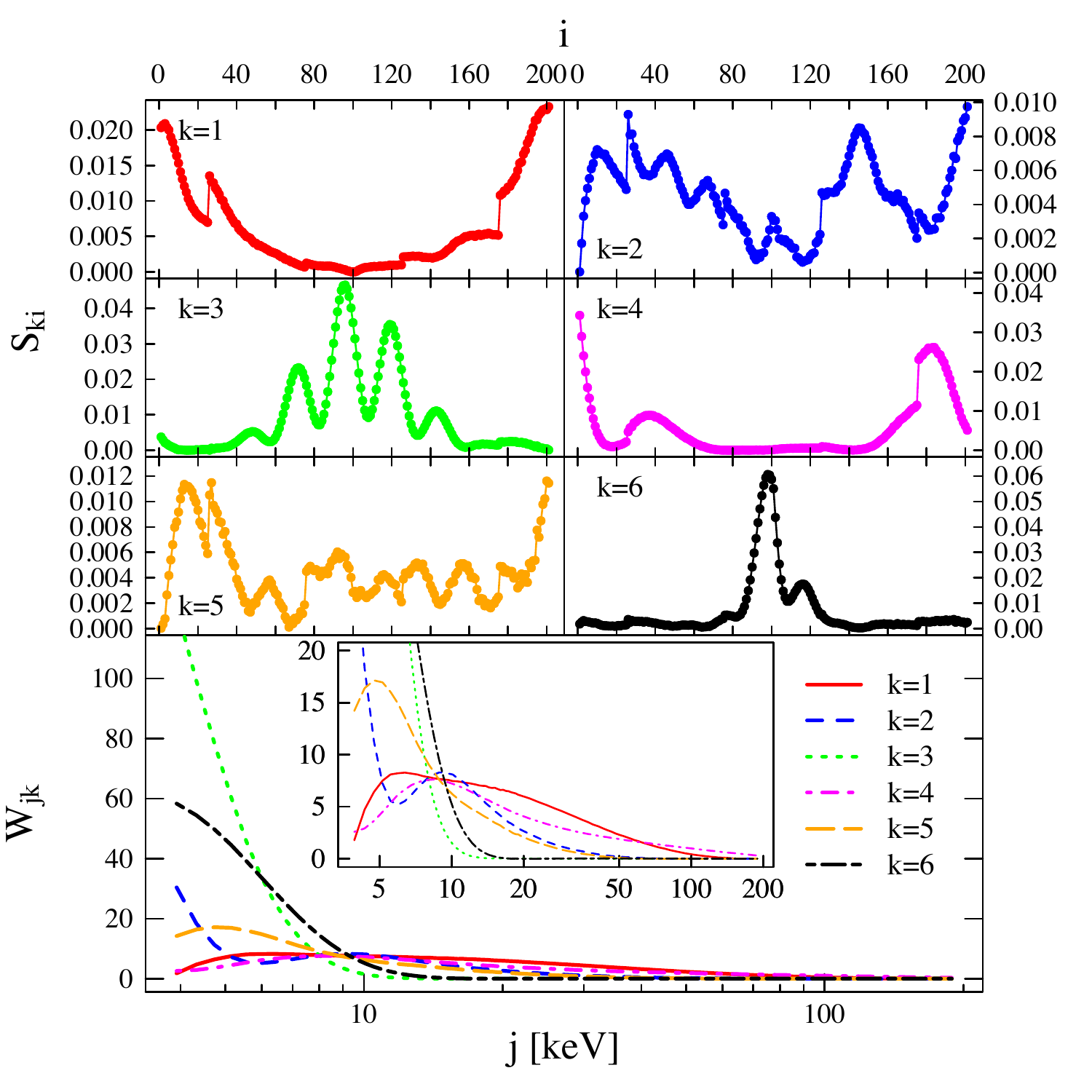}
\end{center}
\vspace{-12pt}
\caption{The six components from the NMF analysis sufficient to explain most of the variability and X-ray spectra of the simulated dataset. The top six panels show the source signals $S_{ki}$ of the NMF components and the bottom panel shows the weights $W_{jk}$ of the NMF components.} \label{nmf}
\end{figure}

Fig. \ref{nmf} shows the components derived by NMF from the simulated dataset. The components are clearly divided to those representing the disc component ($k=3,6$) and the power law component ($k=1,2,4,5$). The two components representing the disc retain even some information of the original parameters, $k=6$ showing the increasing and decreasing disc temperature and $k=3$ sinusoidally varying normalisation, though some mixture of both components can be clearly seen. The components representing the power law are more mixed with each other, but some similarity can be distinguished of the power law normalisation with $k=2$ and $k=5$ components, power law index with $k=1$ component and the cutoff with $k=4$ component. One has to bear in mind that one-to-one correspondence with the model parameters is not expected as their effects to the spectra is non-linear, but NMF performs quite well in distinguishing the disc and power law components from each other and tackling the non-linear effects. In the following we will compare all the decompositions from PCA, ICA and NMF to the fluxes derived from the disc and power law components of the simulated dataset.    

\subsection{The comparison of the decompositions} \label{quality}

It is interesting to note that all methods converge on the same degree of factorisation however with seemingly different components. This shows that individual components of the factorisations do not likely represent any meaningful parameters of the spectral model, which is expected as the parameters vary in non-linear fashion. However, it is possible to compile the disc and power law component from a collection of the factorisation components. Fig. \ref{quality} shows the correlation between the model flux values (see Fig. \ref{sim}) and individual or different combinations of the $S_{ki}$ representing the disc or power law component of the different decomposition methods described above. The PCA can fairly well distinguish the disc flux. The power law flux is trickier to produce, and the $k=2$ component comes closest to that. It is not clear how the correcting components ($k=3,4,5,6$) would improve the match if added or subtracted from these ``main'' components. In addition, it seems that the components do not resemble the variations of the individual model parameters. ICA performs slightly better in distinguishing the flux components. Components $k=3,5,6$ add up to match the disc flux and $k=1,2,4$ to power law flux. However, likewise in PCA the correlation breaks down for low power law fluxes. NMF performs the best in distinguishing the two flux components. The components representing the disc flux ($k=3,6$) are equally good as in PCA and ICA. The components representing the power law flux ($k=1,2,4,5$) distinguish the power law flux of the simulated data the best out of the three methods with smaller spread and continuing the correlation to low power law fluxes as well.   

\begin{figure}
\begin{center}
\includegraphics[width=0.5\textwidth]{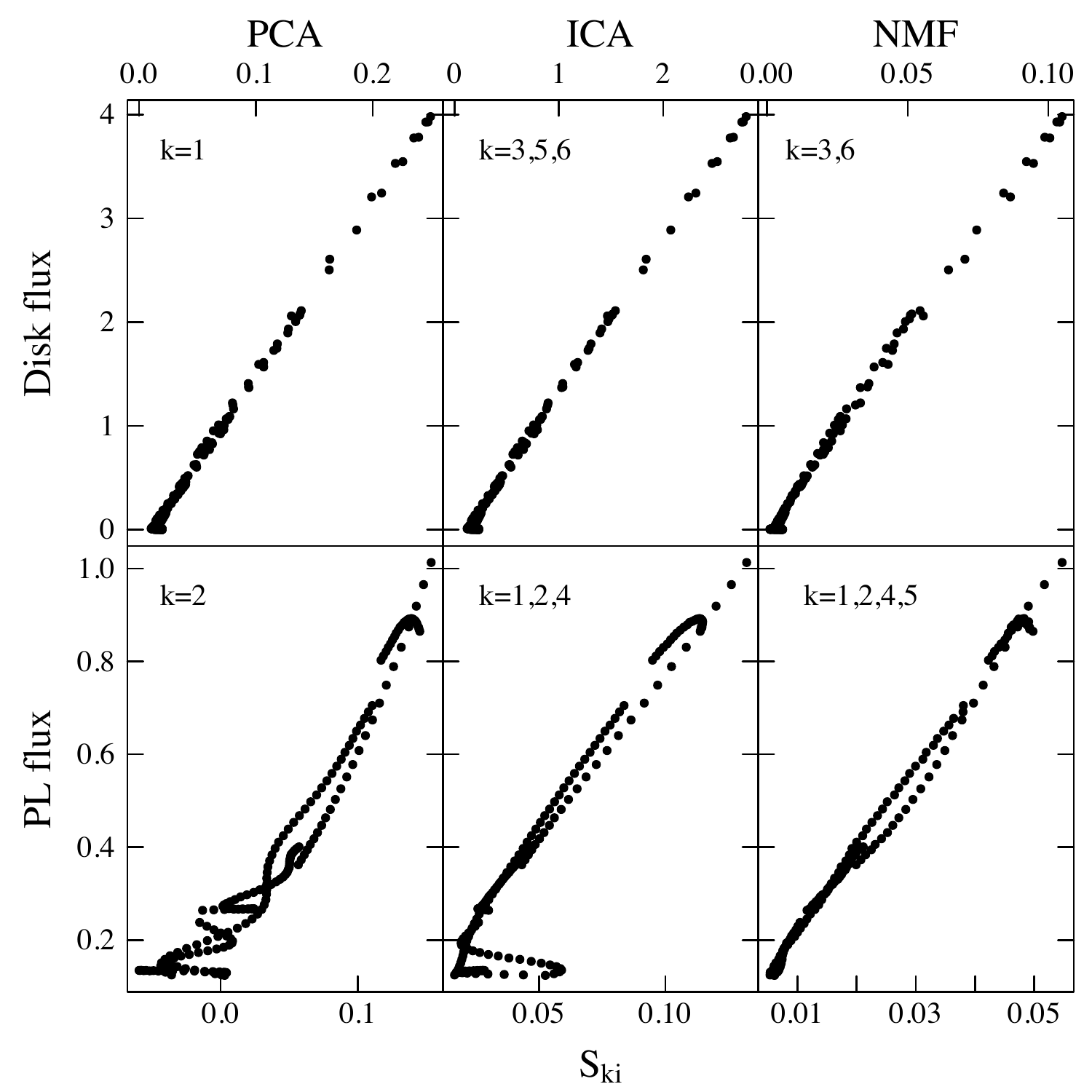}
\end{center}
\vspace{-12pt}
\caption{The quality of the decompositions. The panels show the correlation between the assumed disc and power law fluxes from the PCA (left), ICA (centre) and NMF (right) analyses with the simulated disc (top) and power law (bottom) fluxes (see also Fig. \ref{sim}).} \label{quality}
\end{figure}

\begin{figure}
\begin{center}
\includegraphics[width=0.5\textwidth]{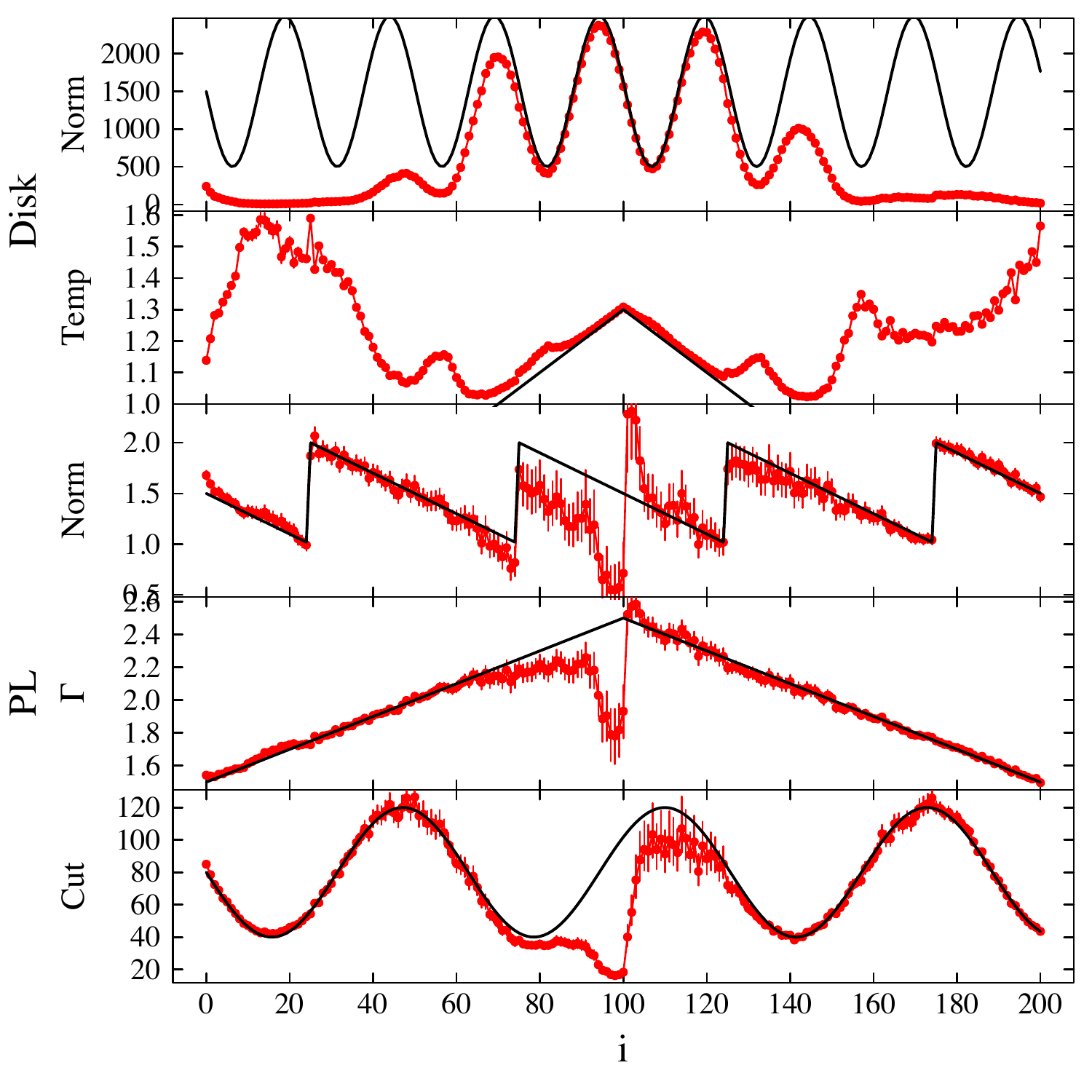}
\end{center}
\vspace{-12pt}
\caption{The spectral parameters (red points) of the NMF spectra fitted independently for disc and power law components as compared to the parameter values of the simulation (solid lines).} \label{plotparams}
\end{figure}

The unique feature of NMF that allows only positive values for the factorisation, allows the resulting $W_{jk}$ and $S_{ki}$ to be united to form the spectra of individual components. Thus, we select the $k$ that corresponds to the disc and power law components mentioned above, and form the disc and power law spectra as $X_{disc}=\sum_{k=3,6}W_{jk}S_{ki}$ and $X_{PL}=\sum_{k=1,2,4,5}W_{jk}S_{ki}$ respectively. The resulting spectra are then fed into \isis\/\footnote{This is done via \isis\/ {\sc load\_data} function, which can load ASCII spectral files as described in the \isis\/ documentation. As the spectra are already flux-corrected, only a diagonal response with 1 cm$^{2}$ area and a nominal 1 sec integration time are assumed for fitting purposes. The units of the ASCII spectra should be photons s$^{-1}$ cm$^{-2}$, thus $X_{disc}$ and $X_{PL}$ are multiplied by the energy bin widths and divided by the bin energy. It is also important to set {\sc minimum\_stat\_err} variable lower than the input uncertainties to keep \isis\/ from modifying them.} and fitted with \diskbb\/ or \cutoffpl\/ model for $X_{disc}$ and $X_{PL}$ respectively. The parameters of the fits are then compared (see Fig. \ref{plotparams}) to the original ones in Fig. \ref{sim}. It is clear that exactly similar parameter tracks are not achieved, but this is mainly driven by low flux values which increase the degeneracy of the model parameters.

\section{Application to GX 339--4} \label{discussion}

GX 339--4 is a low-mass black hole X-ray binary discovered in the early 1970s \citep{markert}. The mass of the black hole is $\geq$ 5.8$M_{\odot}$ \citep{hynes,munoz-darias} at a distance of 7--9 kpc \citep{zdziarski}. The inclination is uncertain though likely less than 60$^\circ$ \citep{cowley}. GX 339--4 exhibits recurrent outbursts and is one of the most observed XRBs, thus making it a very suitable object for detecting temporal changes in its X-ray spectra. In addition, the X-ray spectra of GX 339--4 can be usually fit with relatively simple model consisting of an absorbed multicolour disc black body component and/or a power law component with a cutoff, and an iron line \citep[e.g.][]{dunn}. GX 339--4 is often used as the prime example of a source exhibiting hysteresis in its X-ray spectra, i.e. the change from the power law dominated hard state to the disc dominated soft state occurs at a higher luminosity than the change from the soft state back to the hard state in the same outburst \citep[e.g.][]{miyamoto,nowak95,nowak02}. The outbursts of GX 339--4 also link radio observations to the spectral evolution such that the X-ray flux correlates with the radio flux in the hard state \citep{hannikainen,corbel00,corbel13}, the radio flux exhibits strong, transient outbursts before quenching when the source transits to the soft state, and the radio flux is again detected when the source transits back to the hard state but without strong outbursts \citep{fender}. These characteristics make GX 339--4 a good source to probe accretion physics and the disc-jet connection. Other interesting observations include possible detections of the accretion disc in the hard state \citep{miller, tomsick, reis} which have implications on the accretion geometry and compact jet production.

Thus, we selected GX 339--4 as the source to test the best unsupervised spectral decomposition method found in this paper, namely the NMF analysis (see Section \ref{spectral_decomposition}), because of the amount of data available, the ease of fitting the X-ray spectra with simple models, and the clear hysteresis the source exhibits with implications to the accretion physics.          

\subsection{Observations} \label{observations}

We analysed the archived \rxte\/ observations of GX 339--4 from year 2002 to year 2010 that include both \rxtepca\/ and High-Energy X-ray Timing Experiment (\rxtehexte) data totalling to 934 pointings. We also exclude pointings that exhibit large data gaps in the \rxtehexte\/ spectrum. Small data gaps ($<$3 energy channels) are filled by interpolating data from the neighbouring points. The whole data set includes four outbursts from 2002, 2004, 2007 and 2010. A comprehensive look into these outbursts can be found e.g. from \citet{dunn} and \citet{motta} and references therein. Each pointing is individually reduced by the standard method as described in the \rxte\/ cook book using \heasoft\/ 6.13. The individual spectra for the analysis is prepared as follows: The \rxtepca\/ spectrum is extracted from the top layer of PCU--2, which has been operational for all the selected pointings. The \rxtehexte\/ spectrum is extracted from cluster B for data previous to 14 December 2009 when the cluster stopped rocking and it was permanently left as an off-source pointing position. After that we used cluster A for the source data and estimated the background using cluster B (due to cluster A not rocking and fixed to on-source pointing position). The \rxtepca\/ and \rxtehexte\/ spectra are then unfolded and the flux values extracted from 3--25 keV for \rxtepca\/ and from 20--200 keV for \rxtehexte. \rxtehexte\/ spectra are then normalised to the \rxtepca\/ such that the overlapping energy range from 20--25 keV is the same for both detectors. However, \rxtepca\/ data is used in this region in the final spectra.  

\subsection{Results}

\begin{figure}
\begin{center}
\includegraphics[width=0.5\textwidth]{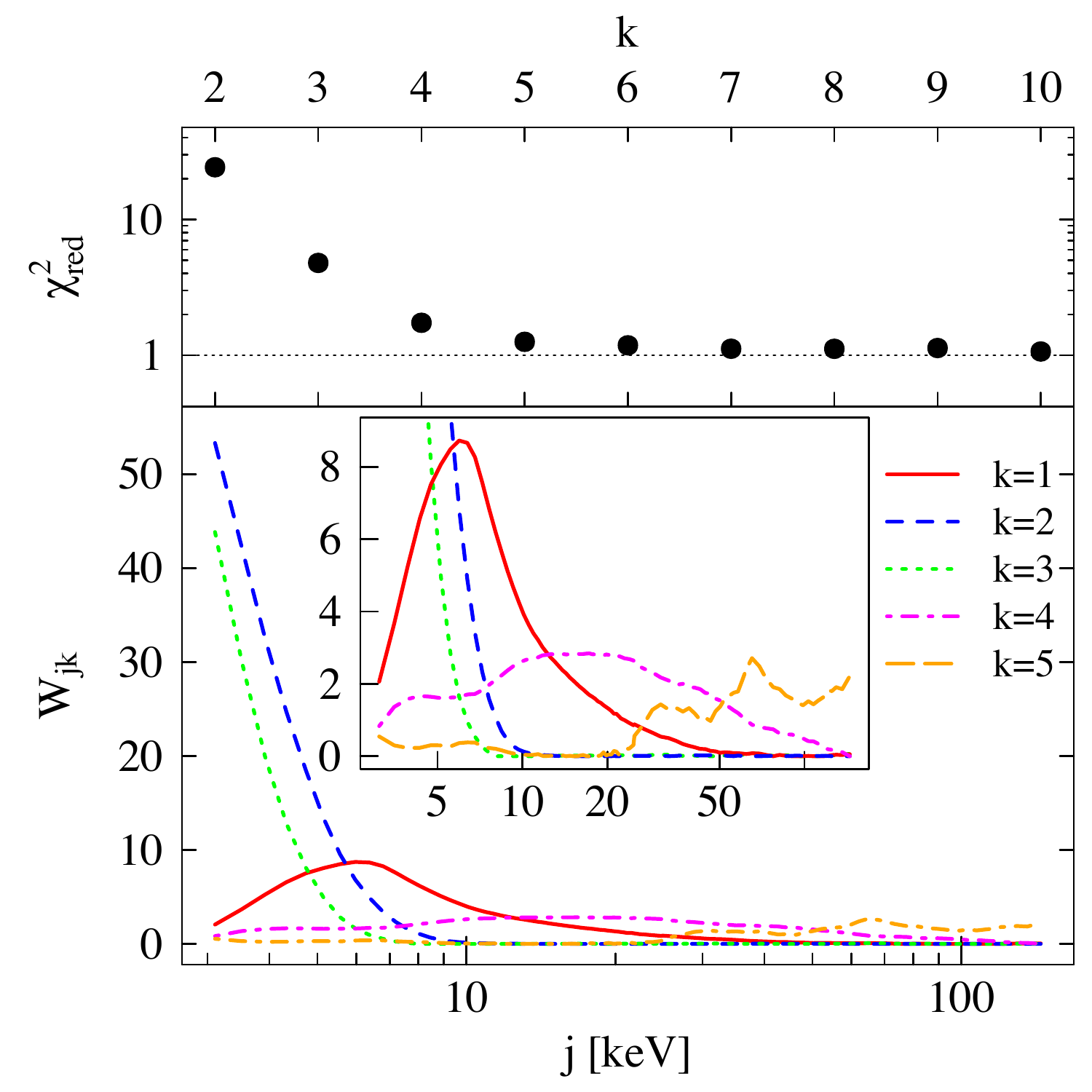}
\end{center}
\vspace{-12pt}
\caption{\textit{Top:} The quality of the degrees of factorisation. We chose $k=5$ for the degree of the NMF analysis (see text for more details). \textit{Bottom:} The individual $W_{jk}$ of the factorisation for each $k$. The inset shows the same plot but for smaller range of $W_{jk}$ to better show components at higher energies. The components for $k=2,3$ can be attributed to the disc, $k=1,4,5$ to the iron line and power law.} \label{plotnmfdata}
\end{figure}  

We used the exact same method as described in Section \ref{nmf_sec} to analyse the spectral data from GX 339--4. Fig. \ref{plotnmfdata} shows the degree of factorisation in the upper panel for GX 339--4 data. We choose $k=5$ for the degree as it provides substantially better approximation than smaller degrees, but only a slightly worse approximation than larger degrees as was explained in Section \ref{nmfdegree}. The lower panel of Fig. \ref{plotnmfdata} shows the individual $W_{jk}$ of the factorisation. From these we can see that the $k=2, 3$ corresponds to the disc, $k=1,4,5$ corresponds to the iron line and the cutoff power law. The $W_{j5}$ component also shows line-like features in the high energies which arise from the \rxtehexte\/ background lines present in the \rxtehexte\/ spectra after 2010. 

As was shown in Section \ref{quality}, we can form the disc and power law fluxes by adding together the appropriate $S_{ki}$, in this case $S_{disc}=\sum_{k=2,3}S_{ki}$ and $S_{PL}=\sum_{k=1,4,5}S_{ki}$ respectively, taking into account the iron line flux in the latter. As was stated in Section \ref{quality} these correspond fairly well to the actual spectral component fluxes. Thus, \textit{it is possible to follow the evolution of the disc and power law fluxes without doing any spectral fitting to the data}. 

In addition, it is possible to construct a disc fraction luminosity diagram (DFLD) and a power law fraction luminosity diagram (PFLD) using $S_{disc}$ and $S_{PL}$. Here we use the PFLD defined as the total flux, $S_{disc}+S_{PL}$, as a function of $S_{PL}/(S_{disc}+S_{PL})$, which equals to 1 when the total flux is composed only of power law flux, and 0 when the total flux is composed only of disk flux. 

\begin{figure*}
\begin{center}
\includegraphics[width=1.0\textwidth]{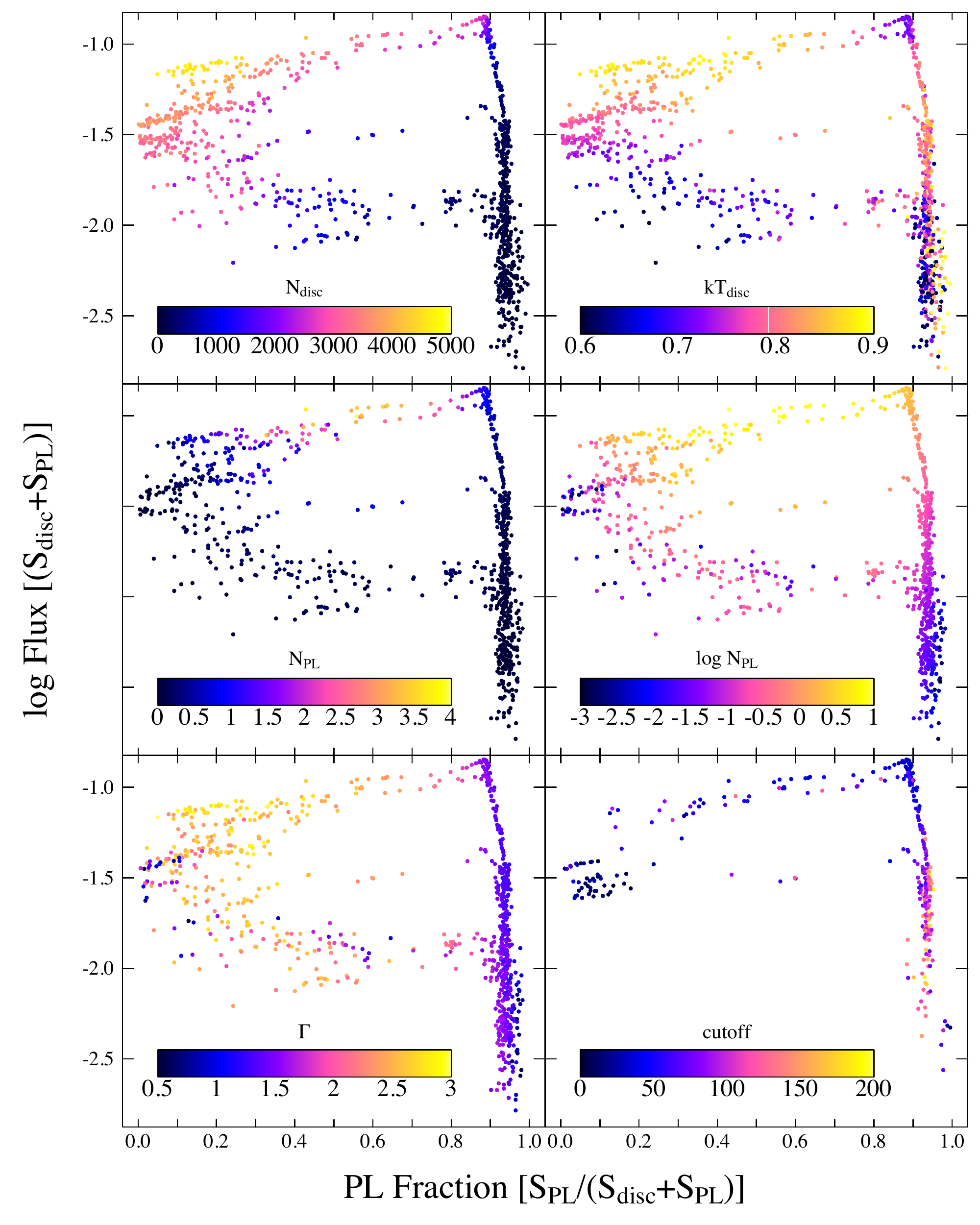}
\end{center}
\vspace{-12pt}
\caption{The change of parameters along the power law fraction luminosity diagram. The upper panels correspond to the disc parameters: normalisation and temperature, the middle and bottom panels correspond to the power law parameters (the power law normalisation is plotted in linear and logarithmic scales in order to distinguish better the parameter evolution for the high and low normalisation regimes).} \label{fitfig}
\end{figure*}

If we take the above results as portraying the actual disc and power law components we can form the disc and power law spectra as $X_{disc}=\sum_{k=2,3}W_{jk}S_{ki}$ and $X_{PL}=\sum_{k=1,4,5}W_{jk}S_{ki}$ respectively, and feed them into \isis\/ for spectral fitting. We use the original data errors for fitting both $X_{disc}$ and $X_{PL}$ spectra and do not add any systematic error to the spectra. The $X_{disc}$ spectra were fit with a model \phabs$\times$\diskbb\/ using an energy range 3--15 keV, and $X_{PL}$ with a model \phabs$\times$\cutoffpl\/ using an energy range 10--140 keV, thus leaving out the iron line region for simplicity and to better concentrate on the power law continuum. Fig. \ref{fitfig} shows the results of the modelling plotted on the PFLD. This is done so as to show the evolution of the parameters rather than to provide hard values. As it is not clear how the data errors would be divided into the $X_{disc}$ and $X_{PL}$ components the resulting parameters have bigger errors than when fitted to the whole spectra. Thus, the parameter values in this presentation are more robust when they are changing gradually, and likely does not represent the actual value when they are in a region where the parameter value changes abruptly across the whole parameter range (i.e. a mix of colours in Fig. \ref{fitfig}). In addition, parameter values that have their error values equal to the minimum or maximum value allowed are removed.

From Fig. \ref{fitfig} we can see the following: 
\begin{enumerate}
\item The power law normalisation increases with the overall luminosity and the power law index increases with the power law fraction, the exception being when the power law fraction is below 0.01. The scatter in the lower part of the hard state appears to be due a changing value of the power law index from $\sim$1--1.5.
\item As the power law fraction starts to decrease in the hard state, the disc normalisation starts to increase. 
\item The cutoff energy decreases with the luminosity in the hard state. Up to the critical luminosity (as discussed above) the cutoff is over 100 keV, but starts to decrease after the critical luminosity down to $\sim$30--40 keV. 
\item The difference between the transition fluxes when transferring to the soft state seem to occur from a difference in the disc normalisation and power law normalisation (an indication of a difference could be seen in the cutoff energy as well), while the disc temperature and power law index change in similar fashion for each transition.  
\item The disc gets hotter when transitioning to the soft state and there is an indication of heating up again when transitioning back to the hard state as well.
\item In the soft state both the disc normalisation and temperature decrease slowly.   
\end{enumerate}

These results correspond well to the ones found in previous studies \citep[e.g. in][]{dunn,delsanto,motta,stiele,cadollebel}. In addition, the NMF analysis is able to track the changes of the disc component in the low disc fraction regime. These correspond to the regions where the disc component is weak and not required statistically in the spectral fits that have been performed previously. However, it is possible that the low energy tail of the response matrices that arise from escape and fluorescence peaks is mimicking the disc component in the faint spectral states. But, these peaks are most likely caused by photons that come from the hard spectral component and thus have the same variability producing soft excess to the hard, and not to the soft, factorised component. This shows the strength of the NMF analysis in situations where the disc component is harder to detect by using only spectral fitting. While the fluxes of the disc and power law components can be distinguished fairly well in the NMF analysis (Fig. \ref{quality}), one has to be careful with the values of spectral components in the low flux regimes as these are most likely degenerate (see Fig. \ref{plotparams} and compare to Fig. \ref{sim}). 

\begin{figure}
\begin{center}
\includegraphics[width=0.5\textwidth]{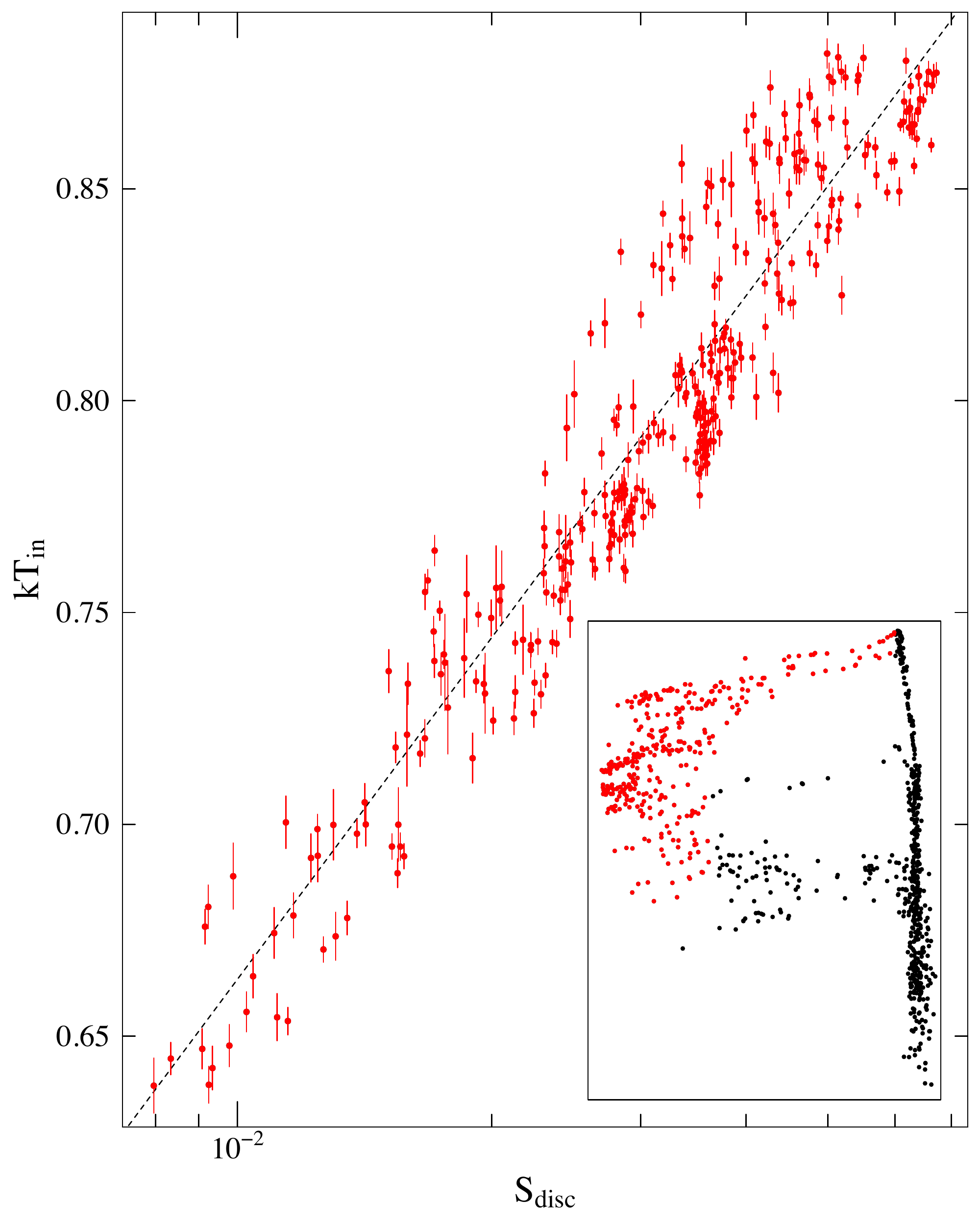}
\end{center}
\vspace{-12pt}
\caption{Temperature-luminosity diagram of a selected set of data points with the best-fit line in black. The inset shows the whole power law fraction luminosity diagram with the selected points marked in red. The best-fit line follows approximately $S_{disc} \sim T^{7}$, which differs from the expected theoretical relation (see text for possible reasons).} \label{lumtemp}
\end{figure}

As was found in previous studies, at the soft state the disc luminosity is driven by the $S_{disc} \sim R_{in}^{2}T^{4}$ -relation mimicking the emission from a black body with radius $R_{in}$ and temperature $T$. It is usually assumed that the above relation is achieved when the inner disc is located at the innermost stable circular orbit (ISCO), though multiple factors can affect this scaling (see a more complete discussion on the difference between these factors in \citealt{salvesen}). Fig. \ref{lumtemp} shows the luminosity-temperature diagram for a set of selected points with a best-fit line in black, and the whole PFLD in the inset with the selected points in red. The best-fit line follows approximately the relation $S_{disc} \sim T^{7}$, which differs from the usual relationship. However, as the \rxte\/ is sensitive only for energies 3 keV and upwards, the missing softer X-ray flux, as well as the absorption, affect the luminosity-temperature diagram as discussed in \citet{dunn} (where their original correlation was $S_{3-10 keV} \sim T^{9.44}$ and corrected $S_{disc} \sim T^{4.75}$). Here we are concerned with the disc-temperature scaling in the broad sense and assume that when taking into account properly the data below 3 keV and the effect of absorption the scaling should be close to $S_{disc} \sim T^{4}$. From Fig. \ref{lumtemp} it can be seen that \textit{GX 339--4 exhibits the disc-temperature relation in the disc component right after it starts the transition to the soft state}, i.e. the inner disc seems to be at the ISCO already in the beginning of the intermediate state right after the peak of the outburst. This relation breaks in the soft state when the power law fraction achieves values close to $\sim$0.3--0.4.  
  
\section{Conclusions} \label{conclusions}

In this paper we have demonstrated that unsupervised linear spectral decomposition methods can be used to follow the evolution of distinct spectral components. The non-linearities present in the original spectral components can be taken into account by adding multiple linear components together based on their weights across the spectral energies. These methods can resolve differently varying spectral components given that they present a measurable effect on the fluxes. Of the three methods used to tackle a simulated data set we showed that the non-negative matrix factorisation performs best over principal component analysis and independent component analysis. The non-negative matrix factorisation also has the additional benefit that the resulting factorisations are always positive, thus the components of these factorisations can be used to fit spectral models separately. Tracking the individual spectral parameter variations is not as robust compared to fluxes but they can be tracked reasonably well with good quality spectra and high flux values. This has the advantage over normal spectral fitting where different models fit the same spectra or a specific spectral component is not required statistically in the fits.

We applied the non-negative matrix factorisation to a set of \rxte\/ spectra of the Galactic black hole X-ray binary GX 339--4 that includes four outbursts from the source. We found that five components are required to accurately represent the spectra, and that these can be divided into disc and power law components taking into account the iron line in the latter. Fitting the factorised disc and power law components separately with an appropriate spectral model corroborates the notion that the inner edge of the disc is at the innermost stable circular orbit at the onset of the intermediate state following the peak of the outburst. However, this analysis should be refined by including data below the \rxte\/ soft X-ray limit. 

In the future we envision that unsupervised linear spectral decomposition methods will be used in multiple situations involving the detection of separate spectral components. In order to extend these preliminary studies in detecting the accretion disc in the low disc flux states, as seen for example in the hard state, we would need to have access to both data from multiple sources and detectors more sensitive to the softer X-ray bands. In order to detect the spectral component responsible for the quasi-periodic oscillations in XRBs, different timing scales could be tested. Finally, these methods provide an alternative way of detecting the spectral components without performing actual spectral fitting, which might prove to be an invaluable tool when dealing with large datasets.    

\section*{Acknowledgements}

I thank the the referee for the careful reading of the manuscript and excellent suggestions that improved the paper. I thank Diana Hannikainen and Thomas Maccarone for useful discussions. I gratefully acknowledge support from Emil Aaltonen s\"a\"ati\"o. This research has made use of data obtained from the High Energy Astrophysics Science Archive Research center (HEASARC), provided by NASA's Goddard Space Flight center.

\label{lastpage}

\end{document}